\documentclass[prb,preprint]{revtex4}

\usepackage{epsf}
\usepackage{epsfig}
\usepackage{textcomp}
\usepackage{dcolumn}
\usepackage{bm}
\usepackage{bbm}
\usepackage{amsfonts}
\usepackage{amsmath}

\begin{document}

\title{Relation between exchange-only optimized potential and 
Kohn-Sham methods with finite basis sets; solution of a paradox}

\author{Andreas G\"orling$^1$, Andreas Hesselmann$^1$, Martin Jones$^2$,
and Mel Levy$^2$}
\affiliation{$^1$Lehrstuhl f\"ur Theoretische Chemie,
Universit\"at Erlangen-N\"urnberg, Egerlandstr. 3, D-91058 Erlangen, Germany}
\affiliation{$^2$ Department of Physics, North Carolina A\&T State University,
  Greensboro, North Carolina 27411, USA}

\date{\today}

\begin{abstract}
  Arguments showing that exchange-only optimized effective potential (xOEP)
  methods, with finite basis sets, cannot in general yield the Hartree-Fock
  (HF) ground state energy, but a higher one, are given. While the orbital
  products of a complete basis are linearly dependent, the HF ground state
  energy can only be obtained via a basis set xOEP scheme in the special case
  that all products of occupied and unoccupied orbitals emerging from the
  employed orbital basis set are linearly independent from each other. In this
  case, however, exchange potentials leading to the HF ground state energy
  exhibit unphysical oscillations and do not represent a Kohn-Sham (KS)
  exchange potential. These findings solve the seemingly paradoxical results
  of Staroverov, Scuseria and Davidson that certain finite basis set xOEP
  calculations lead to the HF ground state energy despite the fact that within
  a real space (or complete basis) representation the xOEP ground state energy
  is always higher than the HF energy.  Moreover, whether or not the occupied
  and unoccupied orbital products are linearly independent, it is shown that
  basis set xOEP methods only represent exact exchange-only (EXX) KS methods,
  i.e., proper density-functional methods, if the orbital basis set and the
  auxiliary basis set representing the exchange potential are balanced to each
  other, i.e., if the orbital basis is comprehensive enough for a given
  auxiliary basis.  Otherwise xOEP methods do not represent EXX KS methods and
  yield unphysical exchange potentials. The question whether a xOEP method
  properly represents a KS method with an exchange potential that is a
  functional derivative of the exchange energy is related to the problem of
  the definition of local multiplicative operators in finite basis
  representations and to the fact that the Hohenberg Kohn theorem does not
  apply in finite basis representations.  Plane wave calculations for bulk
  silicon illustrate the findings of this work.
\end{abstract}

\maketitle

\def\angst{\,\text{\AA}}

\section{\label{sec:introduction}Introduction}
In a recent stimulating article with important implications for the use of
finite basis sets, Staroverov, Scuseria, and Davidson \cite{staroverov06}
presented an exchange-only optimized effective potential (xOEP) scheme that
yields, for given finite Gaussian orbital basis sets, ground state energies
that surprisingly equal exactly the ground state Hartree-Fock (HF) energies
for these basis sets. Moreover, their xOEP scheme not only yields one unique
but an infinite number of exchange potentials and each of the latter leads to
the corresponding ground state HF energy if used as the exchange potential in
the corresponding exchange-only KS Hamiltonian operator. On the other hand, it
is known that in a complete basis set limit, which corresponds to a complete
real space representation of all quantities, the xOEP method is identical
\cite{sahni82} to the exact exchange-only Kohn-Sham method and yields ground
state energies that always lie above \cite{ivanov03} the corresponding ground
state HF energy.  Staroverov, Scuseria, and Davidson then state: "Our
conclusions may appear paradoxical.  For any finite basis set, no matter how
large, there exist infinitely many xOEP’s that deliver exactly the
ground-state HF energy in that basis, however close it may be to the HF limit.
Nonetheless, in the complete basis set limit, the xOEP is unique and E(xOEP)
is above E(HF)". (Here E(xOEP) and E(HF) denote the xOEP and HF total
energies, respectively, that are denoted $E^{xOEP}$ and $E^{HF}$ in this
work.) Furthermore they state: "The non-uniqueness of OEPs in a finite basis
set raises doubt about their usefulness in practical applications"

We here first show, by different means including a constrained-search one,
that the above statement of Staroverov, Scuseria, and Davidson, that it is
always possible to construct optimized effective potentials that deliver
exactly the ground state HF energy, holds if and only if the products of the
orbital basis functions, or at least the products of the corresponding
occupied and unoccupied HF orbitals from a given orbital basis set, form a
linearly independent set. Otherwise, the xOEP scheme for finite orbital basis
sets, in general, does not deliver exactly the ground state HF energy.
Secondly, we show that the xOEP approach of Staroverov, Scuseria, and
Davidson, does not really represent an exchange-only KS method and does not
yield physically meaningfull KS exchange potentials, even if the products of
orbital basis functions are linearly independent.  In order to get physically
meaningfull KS exchange potentials via xOEP schemes, the latter have to be set
up in a way that they represent KS methods, otherwise they are indeed of
little usefulness in practical applications. However, if xOEP schemes are set
up properly then they are of great usefulness in practice as demonstrated,
e.g., by numerically stable plane-wave xOEP procedures for solids
\cite{stadele97,stadele99,magyar04,qteish05,rinke05}.

\section{\label{sec:energies}Relation of {\scriptsize x}OEP  and HF energies
  within finite basis set methods}
We start by briefly reconsidering the xOEP approach of Staroverov, Scuseria,
and Davidson \cite{staroverov06}.  The relevant Hamiltonian operators are the
HF Hamiltonian operator
\begin{eqnarray}
\hat{H}^{HF} &=& -{\textstyle \frac{1}{2}} \nabla^2 
               + {v}_{ext}({\bf r}) 
               + {v}_{H}({\bf r}) + \hat{v}_{x}^{NL} 
\label{HF-Hamiltonian}
\end{eqnarray}
and the exchange-only KS Hamiltonian operator
\begin{eqnarray}
\hat{H}^{xKS} &=& -{\textstyle \frac{1}{2}} \nabla^2 
               + {v}_s({\bf r}) \nonumber \\
              &=& -{\textstyle \frac{1}{2}} \nabla^2 
               + {v}_{ext}({\bf r})
               + {v}_{H}({\bf r}) + {v}_{x}({\bf r}) 
\label{KS-Hamiltonian}
\end{eqnarray}
Atomic units are used throughout.  In Eqs. (\ref{HF-Hamiltonian}) and
(\ref{KS-Hamiltonian}), ${v}_{ext}({\bf r})$ denotes the external potential,
usually the electrostatic potential of the nuclei, ${v}_{H}({\bf r})$ is the
Hartree potential, i.e., the Coulomb potential of the electron density,
${v}_{x}({\bf r})$ is the local multiplicative KS exchange potential,
${v}_s({\bf r})= {v}_{H}({\bf r}) + {v}_{x}({\bf r}) + {v}_{ext}({\bf r})$ the
effective KS potential, and $\hat{v}_{x}^{NL}$ the nonlocal exchange operator
with the kernel
\begin{eqnarray}
\hat{v}_{x}^{NL}({\bf r},{\bf r}')  &=& 
\frac{\rho({\bf r},{\bf r}')}{|{\bf r}-{\bf r}'|}
\label{HF-kernel}
\end{eqnarray}
in a real space representation. Here $\rho({\bf r},{\bf r}')$ designates the
first-order density matrix. In the HF-Hamiltonian operator of Eq.\
(\ref{HF-Hamiltonian}) the first-order density matrix occuring in the nonlocal
exchange operator of Eq. (\ref{HF-kernel}) equals the HF first order density
matrix $\rho^{HF}({\bf r},{\bf r}')$ and the nonlocal exchange operator
subsequently equals the HF exchange operator.  For simplicity we consider
closed shell systems with non-degenerate ground states. In this case orbitals,
first-order density matrices, and basis functions can all be chosen to be
real-valued.

Next we introduce an orbital basis set $\{\chi_{\mu}\}$ of dimension $N$.  The
representations of the HF- and exchange-only KS-Hamiltonian operators in this
basis set are
\begin{eqnarray}
{\bf H}^{HF} &=& {\bf T} + {\bf V}_{H} + {\bf V}_{x}^{NL} + {\bf V}_{ext} 
\label{HF-Hamiltonian-basis}
\end{eqnarray}
and 
\begin{eqnarray}
{\bf H}^{xKS} &=& {\bf T} + {\bf V}_s \nonumber \\
              &=& {\bf T} + {\bf V}_{H} + {\bf V}_{x} + {\bf V}_{ext} \,,
\label{KS-Hamiltonian-basis}
\end{eqnarray}
respectively. The matrices ${\bf T}$, ${\bf V}_{H}$, ${\bf V}_{ext}$, 
${\bf V}_{x}^{NL}$ and ${\bf V}_{x}$ are defined by the corresponding matrix
elements $T_{\mu\nu} 
= \langle\chi_{\mu} | -{\textstyle \frac{1}{2}} \nabla^2 | \chi_{\nu}\rangle$,
$V_{H,\mu\nu} 
= \langle\chi_{\mu} | {v}_{H} | \chi_{\nu}\rangle$,
$V_{ext,\mu\nu} 
= \langle\chi_{\mu} | {v}_{ext} | \chi_{\nu}\rangle$,
$V_{x,\mu\nu}^{NL} 
= \langle\chi_{\mu} | \hat{v}_{x}^{NL} | \chi_{\nu}\rangle$, and
$V_{x,\mu\nu} 
= \langle\chi_{\mu} | \hat{v}_{x} | \chi_{\nu}\rangle$, respectively, and by
${\bf V}_s = {\bf V}_{H} + {\bf V}_{x} + {\bf V}_{ext}$. Because the orbital
basis functions are real-valued all matrices are symmetric

Now we expand the KS exchange potential in an auxiliary basis set
$\{f_{k}\}$ of dimension $M^{aux}$, i.e.,
\begin{eqnarray}
{v}_{x}({\bf r}) = \sum_{k=1}^{M^{aux}} \; b_k \, f_{k}({\bf r}) \,.
\label{vx-aux}
\end{eqnarray}
The auxiliary basis set, of course, shall be chosen such that its basis
functions are linearly independent. The crucial question arising now is how
many and what types of matrices ${\bf V}_{x}$ representing the KS
density-functional exchange potential can be constructed for a given auxiliary
basis set $\{f_{k}\}$. This question was answered in Ref.
[\onlinecite{gorling95}]. Firstly we consider the case when the $M = (1/2) N
(N+1)$ different products $\chi_{\mu}({\bf r}) \chi_{\nu}({\bf r})$ of orbital
basis functions are linearly independent. In this case, if $M^{aux} = M$ and
the auxiliary basis functions span the same space as the products of the
orbital basis functions, then a symmetric matrix ${\bf V}_{x}$ can be
constructed in a unique way by determining appropriate expansion coefficients
$b_k$ for the exchange potential. The reason is that the determination of the
$M^{aux} = M$ expansion coefficients $b_k$ for the construction of the
$M^{aux} = M$ different matrix elements of the symmetric matrix ${\bf V}_{x}$
leads to a linear system of equations
\begin{eqnarray}
{\bf A} \, {\bf b}  = {\bf W}_{x}
\label{equationforXcoefficients}
\end{eqnarray}
with
\begin{eqnarray}
{A}_{\mu\nu,t}  = \langle\chi_{\mu}   \chi_{\nu} | f_k\rangle
\label{A}
\end{eqnarray}
for the coefficients $b_k$ of dimension $M^{aux} = M$ that is nonsingular and
thus has a unique solution \cite{gorling95}. In Eq. \
(\ref{equationforXcoefficients}), ${\bf A}$ is a $M \times M^{aux}$ matrix
that contains the overlap matrix elements $\langle\chi_{\mu} \chi_{\nu} |
f_k\rangle$. The first index of ${\bf A}$, i.e., $\mu\nu$, is a superindex
refering to products of orbital basis functions while the second index, $k$,
refers to auxiliary basis functions. The vector ${\bf b}$ collects the
expansion coefficients of Eq.\ (\ref{vx-aux}) for the exchange potential and
the right hand side ${\bf W}_{x}$, a vector with superindices $\mu\nu$,
contains the $M = N(N+1)/2$ independent elements of an arbitrarily chosen
matrix ${\bf V}_{x}$. If we chose ${\bf V}_{x}$ to be equal to the matrix
representation of an arbitrary nonlocal operator with respect to the orbital
basis set then Eqs.\ (\ref{vx-aux}) and (\ref{equationforXcoefficients})
define a local potential with the same matrix representation. This
demonstrates that a distinction of local multiplicative and nonlocal operators
is not clearly possible for orbital basis sets with linearly independent
products of orbital basis functions.

If $M^{aux} > M$ and the space spanned by the auxiliary functions contains the
space spanned by the product of orbital functions then \cite{gorling95} an
infinite number of sets of coefficients $b_k$ lead to any given symmetric
matrix ${\bf V}_{x}$.  The real space KS exchange potentials ${v}_{x}({\bf
  r})$ corresponding according to Eq.\ (\ref{vx-aux}) to these sets of
coefficients $b_k$ are all different but all represent local multiplicative
potentials. Next we construct KS Hamiltonian operators (\ref{KS-Hamiltonian})
by adding these different KS exchange potentials to always the same external
and Hartree potential. The resulting effective KS potentials in real space,
i.e., the ${v}_s({\bf r})$ are all different. Nevertheless the resulting basis
set representations ${\bf H}^{xKS}$ of the corresponding KS Hamiltonian
operators are all identical because the basis set representations ${\bf
  V}_{x}$ of the different exchange potentials ${v}_{x}({\bf r})$, by
construction, are all identical.  As a consequence the KS orbitals resulting
from diagonalizing the KS Hamiltonian matrix ${\bf H}^{xKS}$ and subsequently
also the resulting ground state electron densities are identical in all cases.
We thus have a situation where different local multiplicative KS potentials
${v}_s({\bf r})$ lead to the same ground state electron density. This seems to
constitute a violation of the Hohenberg-Kohn theorem.  Indeed it was shown in
Ref. [\onlinecite{harriman86}] and discussed in Ref. [\onlinecite{gorling95}]
that the Hohenberg-Kohn theorem does not hold for finite orbital basis sets in
its original formulation, i.e., that different local potentials, e.g., local
potentials obtained by different linear combinations of auxiliary basis
funtions, must lead to different KS determinants and thus different KS
electron densities. We will come back to this point later on.  Finally, if
$M^{aux} < M$ then not all symmetric matrices ${\bf V}_{x}$ can be constructed
from a local KS exchange potential given by an expansion (\ref{vx-aux}).

In their xOEP approach Staroverov, Scuseria, and Davidson \cite{staroverov06}
can expand the KS exchange potential in $M^{aux} = M$ auxiliary basis
functions and determine the coefficients such that the resulting matrix ${\bf
  V}_{x}$ exactly equals the HF exchange matrix ${\bf V}_{x}^{NL}$. If
additionally the KS Hartree potential is set equal to the HF one then the
resulting HF and KS Hamiltonian operators are identical. Subsequently also the
HF and KS orbitals, the ground state electron densities, and the ground state
energies are identical. Because the HF and the KS electron densities turn out
to be identical, the Coulomb potential of this density can equally well be
considered as a HF or a KS Hartree potential. It follows immediately that the
KS exchange potential constructed in this way is the xOEP exchange potential:
The HF total energy is the lowest total energy any Slater determinant can
yield. Thus if a local multiplicative KS potential leads to this total energy
it is clearly the optimized effective potential defined as the potential that
yields the lowest total energy achievable by any local multiplicative KS
potential. The xOEP ground state energy resulting from this construction
equals the corresponding HF energy. Moreover by enlarging the number of
auxiliary basis functions, resulting in $M^{aux} > M$, not only one optimized
exchange potential leading to the HF energy but infinitely many can be
constructed.

Staroverov, Scuseria, and Davidson obtained the HF energy in their xOEP scheme
even if the number of auxiliary functions only equaled the product $M^{ov}$ of
occupied and virtual orbitals \cite{staroverov06}. In this case a similarity
transformation of the HF and the KS Hamiltonian matrices and their
constituents was carried out in order to obtain representations of all
matrices with respect to the HF orbitals.  Then it is sufficient to chose the
expansion coefficients of the KS exchange potential such that only the
occupied-virtual block of the KS exchange matrix equals that of the HF
exchange matrix.  The resulting KS Hamiltonian matrix then may differ from the
HF Hamiltonian matrix in the occupied-occupied and the virtual-virtual block
but this merely leads to unitary transformations of the occupied and virtual
orbitals among themselves and thus does not change the ground state energy or
the electron density.

Indeed it is straightforward to show that the occupied-virtual block of the
exchange matrix equals that of the HF exchange matrix if the products of
occupied and unoccupied orbitals are linearly independent.  To that end we
consider the xOEP equation determining the xOEP exchange potential
\cite{sharp53,talman76}
\begin{eqnarray}
4 \sum_i^{occ.}\sum_a^{unocc.} \;
\phi_i({\bf r})\phi_a({\bf r}) \; 
\frac{\langle \phi_a | v_x | \phi_i \rangle}
{\varepsilon_i - \varepsilon_a} 
&=& 
4 \sum_i^{occ.}\sum_a^{unocc.} \;
\phi_i({\bf r})\phi_a({\bf r}) \; 
\frac{\langle \phi_a | v_x^{NL}| \phi_i \rangle}
{\varepsilon_i - \varepsilon_a} \,.
\label{EXXb}
\end{eqnarray}
In Eq.\ (\ref{EXXb}) $\phi_i$ and $\phi_a$ denote occupied and unoccupied KS
orbitals, respectively, with eigenvalues $\varepsilon_i$ and $\varepsilon_a$.
Both sides of Eq.\ (\ref{EXXb}) are a linear combination of products
$\phi_i({\bf r})\phi_a({\bf r})$ of occupied and unoccupied KS orbitals with
coefficients ${\langle \phi_a | v_x | \phi_i \rangle}/{(\varepsilon_i-
  \varepsilon_a)}$ and ${\langle \phi_a | v_x^{NL}| \phi_i
  \rangle}/{(\varepsilon_i- \varepsilon_a)}$, respectively. However, if the
products $\phi_i({\bf r})\phi_a({\bf r})$ are linearly independent then the
two linear combinations can only be identical if the coefficients multiplying
the products are all identical. This, however, requires that $\langle \phi_a |
v_x | \phi_i \rangle = \langle \phi_a | v_x^{NL} | \phi_i \rangle$, i.e., that
the occupied-virtual block of the KS exchange matrix equals that of the
corresponding exchange matrix of a nonlocal exchange operator of the form of
the HF exchange operator. Replacement of the KS exchange matrix by the matrix
of the nonlocal exchange operator thus again leads only to a unitary
transformation of the occupied and virtual orbitals among themselves.
Therefore the corresponding xOEP determinant can also be interpreted as HF
determinant.

Next we consider the crucial point what happens if the products of orbital
basis functions $\chi_{\mu}({\bf r}) \chi_{\nu}({\bf r})$ are linearly
dependent. Then the rows of the matrix ${\bf A}$ of Eq.\
(\ref{equationforXcoefficients}) are linearly dependent, thus the rank of the
matrix ${\bf A}$ is lower than $M$, and as consequence Eq.\
(\ref{equationforXcoefficients}), in general, has no solution. For an
alternative argument, observe that for linear dependent products of orbital
basis functions $\chi_{\mu}({\bf r}) \chi_{\nu}({\bf r})$, there exists at
least one linear combination of such products that equals zero
\begin{eqnarray}
0 = \sum_{\mu\nu} \; a_{\mu\nu} \, \chi_{\mu}({\bf r}) \chi_{\nu}({\bf r}) 
\,.
\label{linear-dependence}
\end{eqnarray}
In Eq.\ (\ref{linear-dependence}) the $a_{\mu\nu}$ denote the coefficients of
that linear combination.  The corresponding sum of matrix elements of ${\bf
  V}_{x}$ also equals zero, i.e.,
\begin{eqnarray}
0 = \sum_{\mu\nu} \; a_{\mu\nu} \, \langle \chi_{\mu} | v_x| \chi_{\nu}\rangle
 = \int \! d{\bf r} \, v_x({\bf r}) \,
\sum_{\mu\nu} \; a_{\mu\nu} \, \chi_{\mu}({\bf r}) \chi_{\nu}({\bf r}) 
\label{zero-matrix-elements}
\end{eqnarray}
for any choice of expansion coefficients $b_t$ in Eq.\ (\ref{vx-aux}) because
the product of any local function and thus of any KS exchange potential
${v}_{x}({\bf r})$ with the sum (\ref{linear-dependence}) equals zero. The
products $\chi_{\mu}({\bf r}) \chi_{\nu}({\bf r}')$ for two different
arguments ${\bf r}$ and ${\bf r}'$, on the other hand, are always linearly
independent because the orbital basis set $\{\chi_{\mu}\}$ has to be linearly
independent. Therefore the linear combination $\sum_{\mu\nu} \; a_{\mu\nu} \,
\chi_{\mu}({\bf r}) \chi_{\nu}({\bf r}')$ can not be identical to zero for all
values of the arguments ${\bf r}$ and ${\bf r}'$. Then, however, also the
integral of this linear combination with $\rho^{HF}({\bf r},{\bf r}')/|{\bf
  r}-{\bf r}'|$, i.e., with the kernel of the nonlocal HF exchange operator,
in general, is not equal to zero, i.e., in general
\begin{eqnarray}
0 \ne \sum_{\mu\nu} \; a_{\mu\nu} \, 
\langle \chi_{\mu} | \hat{v}_x^{HF}| \chi_{\nu}\rangle \,.
\label{zero-matrix-elementsHF}
\end{eqnarray}
Comparison of Eqs. (\ref{zero-matrix-elements}) and
(\ref{zero-matrix-elementsHF}) shows that, in general, the exchange matrices
${\bf V}_{x}$ and ${\bf V}_{x}^{NL}$ are different no matter how the expansion
coefficients $b_k$ of the KS exchange potential, Eq. (\ref{vx-aux}), are
chosen. This demonstrates that, in general, neither the xOEP scheme of
Staroverov, Scuseria, and Davidson \cite{staroverov06} nor any other leads to
an xOEP Hamiltonian operator that equals the HF Hamiltonian operator when the
orbital basis products are linearly dependent.  If we consider the version of
Staroverov, Scuseria, and Davidson's xOEP scheme that refers only to the
occupied-virtual block of the xOEP and HF exchange matrices then by completely
analogous arguments it follows that this scheme only works if the products of
occupied and unoccupied HF orbitals are linearly independent. However, in
general, if the products of occupied and unoccupied HF orbitals are linearly
dependent then it is not possible to obtain the HF ground state energy via an
xOEP scheme.

\subsection{\label{subsec:constrainedsearch} Constrained-search analysis}
Before we discuss the question how products of basis functions can become
linearly dependent for given orbital basis sets we elucidate the situation
from a constrained-search \cite{levy79} point of view.  We start with a
constrained-search proof that the xOEP ground state energy, $E^{xOEP}$, must
equal the HF ground state energy, $E^{HF}$, in their common finite orbital
basis, when there is no linear dependence in the products of orbital basis
functions.  To accomplish this we appeal to the work of Harriman
\cite{harriman86}.  He showed that only one first-order density matrix may
yield any density generated by a given finite orbital basis whose basis
products form a linearly independent set.  This means that since an idempotent
first-order density matrix uniquely fixes a corresponding single determinant,
it follows that only one single determinant, constructed from a given finite
orbital basis whose products are linearly independent, may yield a density
that is constructed from this same basis.  Consequently, with use of a common
finite orbital basis set, the xOEP single determinant must equal the HF single
determinant if there exists an effective KS potential $v_s$ in Eq
(\ref{KS-Hamiltonian}) such that the corresponding KS ground state density is
the same as the Hartree-Fock density.  That this $v_s$ exists for the
situation when the basis products are linearly independent, as discussed
above, follows from Ref.[\onlinecite{gorling95}] and was shown in practice by
Staroverov, Scuseria, and Davidson \cite{staroverov06}.

What happens when the products are not linearly independent?  Due to the
idempotency property of the first-order density matrix for a single
determinant, a density generated from a given finite orbital basis could still
generate a unique determinant if the basis products are linearly dependent,
provided that this linear dependency is mild enough \cite{levy87}, i.e., if
the products of occupied and unoccupied orbitals remain linearly independent.
However, if the linear dependency of the basis product pairs is not
sufficiently mild, then the situation changes dramatically in that more than
one single determinant will yield the same density from a given finite basis
set \cite{levy87}.  In this case we do not have equality $E^{xOEP}=E^{HF}$.
Instead, we have inequality $E^{xOEP} > E^{HF}$, which arises from the
following contradiction.

Assume that the xOEP determinant $\Phi^{xOEP}$ equals the HF determinant
$\Phi^{HF}$ through respective optimizations in their common finite orbital
basis set.  Then it follows that their densities must be the same.  But, from
a constrained-search analysis \cite{constrainedsearch}, the xOEP determinant
$\Phi^{xOEP}$ would yield this HF density and minimize, within this common
basis, just the expectation value $\langle\Phi | \hat{T} |\Phi\rangle$ of the
kinetic energy, while the HF determinant $\Phi^{HF}$ yields this HF density
and minimizes, within the common basis, the expectation value $\langle\Phi |
\hat{T} + \hat{V}_{ee} |\Phi\rangle$ of the kinetic energy plus the
electron-electron repulsion energy. Here $\hat{T}$ denotes the many-electron
kinetic energy operator, $\hat{V}_{ee}$ the corresponding electron-electron
repulsion operator, and $\Phi$ Slater determinants that yield the HF density.
(Equivalently, the xOEP determinant would yield the HF density and minimize
$\langle\Phi | \hat{H}-\hat{V}_{ee} | \Phi\rangle$ while the HF determinant
yields this HF density and of course minimizes $\langle\Phi |\hat{H}
\Phi\rangle$.  Here $\hat{H}$ denotes the many-electron Hamiltonian
operator.).  Because the Slater determinants $\Phi^{xOEP}$ and $\Phi^{HF}$
minimize different expectation values, i.e., $\langle\Phi | \hat{T}
|\Phi\rangle$ and $\langle\Phi | \hat{T} + \hat{V}_{ee} |\Phi\rangle$,
respectively, they are different, in general, and the inequality $E^{xOEP} >
E^{HF}$ applies for this common finite orbital basis case.
However, there is only one possible determinant $\Phi$ that yields the HF
density from a given finite basis when the basis products are linearly
independent or the extent of linear dependency is weak. In this case both
minimizations yield this one Slater determinant simply because both
minimization only run over one Slater determinant. Thus there is no
contradiction and the finite basis set conclusion of Staroverov, Scuseria, and
Davidson follows in that the equality $E^{xOEP}=E^{HF}$ applies.  Hence we are
now able to provide the resolution of the xOEP paradox \cite{staroverov06}
stated by Staroverov, Scuseria, and Davidson: For a finite basis set case, no
matter how large the basis, $E^{xOEP}$ equals $E^{HF}$ provided that the basis
products form a linearly independent set or the extent of linear dependence is
sufficiently weak.  However, in going from any starting finite basis set to
the complete basis set limit, $E^{xOEP}$ may become greater than $E^{HF}$
somewhere along the way because as more and more basis orbitals are added to
the finite basis set, the onset of sufficient linear dependency eventually
occurs (see Appendix) .

We have provided an explanation for what might very well seem counterintuitive
to the reader without knowledge of the analysis provided here. As one keeps
adding more and more orbital basis functions, both $E^{HF}$ and $E^{xOEP}$
decrease and they continually remain equal to each other. Past a certain
΄critical point‘ in the addition of orbital basis functions, however, $E^{HF}$
and $E^{xOEP}$ may start to differ from each other and $E^{HF}$ keeps
decreasing while the behavior of $E^{xOEP}$ depends on the chosen orbital
basis set and it might actually be that $E^{xOEP}$ rises!  The latter behavior
for example occurs if the exact HF orbitals as they correspond to a real space
representation are themselves chosen as the basis set.  If the basis set is
restricted to the occupied HF orbitals, $E^{HF}$ and $E^{xOEP}$ are of course
equal. If unoccupied HF orbitals are added to the basis set, $E^{HF}$ remains
unchanged at first.  In contrast, beyond a certain point $E^{xOEP}$ raises.
The cause, of course, is the appearance of sufficient linear dependence at the
critical point. (Ref. [\onlinecite{staroverov06}] does analyse certain linear
dependency situations but the authors do not discuss the energy consequences
for finite basis sets.)

\subsection{\label{subsec:lineardependence} Creation of linear dependence}
Next we consider how products of orbitals basis functions become linearly
dependent.  As example we consider a plane wave basis set corresponding to a
unit cell defined by the three linearly independent lattice vectors ${\bf
  a}_1$, ${\bf a}_2$, and ${\bf a}_3$. The plane waves representing the
orbital basis set $\{\chi_{\bf G}\}$ then are given by
\begin{eqnarray}
\chi_{\bf G}({\bf r}) = \frac{1}{\sqrt{\Omega}} \; e^{i{\bf G}{\bf r}}
\label{plane-waves}
\end{eqnarray}
with 
\begin{eqnarray}
{\bf G} = \ell \, {\bf b}_1 + m \, {\bf b}_2 + n \, {\bf b}_3 
\label{G}
\end{eqnarray}
and 
\begin{eqnarray}
\ell, n , m  \; \in \; \mathbbm{Z}  \;\;\;\;\;\; \mbox{and} \;\;\;\;\;\;
|{\bf G}| \le {G}_{cut} \,.
\label{ellmn}
\end{eqnarray}
In Eq.\ (\ref{G}), ${\bf b}_1$, ${\bf b}_2$, ${\bf b}_3$ denote three
reciprocal lattice vectors defined by the conditions ${\bf a}_{\ell} \cdot
{\bf b}_{m} = 2 \pi \delta_{\ell m}$ for $\ell, m = 1,2,3$. By $\mathbbm{Z}$
the space of all integer numbers is denoted, ${G}_{cut}$ denotes the cutoff
that determines the size of the plane wave basis set, and $\Omega$ stands for
the crystal volume.  We have assumed before that basis functions are
real-valued.  This is not the case for plane waves. However, we can always
obtain a real valued basis set by linear combining all pairs of plane waves
with wave vectors ${\bf G}$ and $-{\bf G}$ to real-valued basis functions.
This real valued basis set and the original complex-valued plane wave basis
set are related by a unitary transformation that does not change any of the
arguments of this paper. All arguments therefore are also valid for the
complex-valued plane wave basis sets considered here and below.  The number
$M$ of basis functions roughly equals $(4\pi/3){G}_{cut}^3 (V/8\pi^3)$. The
exact value of $M$ depends on whether reciprocal lattice vectors ${\bf G}$
that lie in the immediate vicinity of the surface of the sphere with radius
${G}_{cut}$ have lengths that are slightly larger or slightly smaller than
${G}_{cut}$.  The relation
\begin{eqnarray}
\chi_{\bf G}({\bf r}) \chi_{{\bf G}'}({\bf r}) 
= V^{-1} \; e^{i{\bf G}{\bf r}} e^{i{\bf G}'{\bf r}}
= V^{-1} \; e^{i({\bf G}+{\bf G}'){\bf r}} = 
\frac{1}{\sqrt{V}}  \; \chi_{{\bf G}+{\bf G}'}({\bf r})
\label{plane-waves-products}
\end{eqnarray}
shows that the products of plane waves of the orbital basis set are again
plane waves of the same type with reciprocal lattice vectors ${\bf G}+{\bf
  G}'$ that obey the relation ${\bf G}+{\bf G}'\le 2 {\bf G}_{cut}$. Due to
the latter relation the number of different products $\chi_{\bf G} \chi_{{\bf
    G}'}$ is about 8 times as large as the number $N$ of orbital basis
functions, i.e., equals about $8N$.  If $N > 15$ then $8 N < N (N+1)/2$. In
this case the number of {\em different} products of orbital basis functions is
smaller than the number of products of orbital functions. Thus some products
of orbital functions are equal and thus linearly dependent. For realistic
systems the number of plane wave basis functions is much larger than 15. In a
plane wave framework therefore xOEP and HF methods, in general, lead to
different ground state energies with $E^{xOEP} > E^{HF}$. Results from plane
wave xOEP and HF calculations for silicon discussed below illustrate this
point.

\section{\label{sec:potentials}Relation of 
{\scriptsize x}OEP and exchange-only KS methods}
In this Section we show that the xOEP approach of Staroverov, Scuseria, and
Davidson \cite{staroverov06} does not really correspond to an exact exchange
KS method and does not yield a KS exchange potential, irrespective of whether
or not the products of basis functions of the chosen orbital basis set are
linearly independent.  To this end we consider the xOEP or exact exchange
(EXX) equation written in a form that slightly differs from that of Eq.\
(\ref{EXXb})
\begin{eqnarray}
\int \! d{\bf r}' \; X_s({\bf r},{\bf r}') \, v_x({\bf r}')  
&=& 
4 \sum_i^{occ.}\sum_a^{unocc.} \;
\phi_i({\bf r})\phi_a({\bf r}) \; 
\frac{\langle \phi_a | v_x^{NL}| \phi_i \rangle}
{\varepsilon_i- \varepsilon_a} \,.
\label{EXX}
\end{eqnarray}
The response function $X_s$ in Eq.\ (\ref{EXX}) is given by
\begin{eqnarray}
X_s({\bf r},{\bf r}') &=& 
4 \sum_i^{occ.}\sum_a^{unocc.} \;
\frac{\phi_i({\bf r})\phi_a({\bf r})\phi_a({\bf r}')\phi_i({\bf r}')}
{\varepsilon_i- \varepsilon_a}\,.
\label{Xs}
\end{eqnarray}
Eq.\ (\ref{EXX}) can be derived in completely different ways.  Firstly,
following Refs. [\onlinecite{sharp53}] and [\onlinecite{talman76}], one can
consider the expression of the HF total energy and search for those orbitals
that minimize this energy under the constraint that the orbitals are
eigenstates of a Schr\"odinger equation with an Hamiltonian operator of the
form
\begin{eqnarray}
\hat{H}^{OEP} &=& -{\textstyle \frac{1}{2}} \nabla^2 
               + {v}^{OEP}({\bf r}) \,.
\label{OEP-Hamiltonian}
\end{eqnarray}
The search for these orbitals is tantamount to searching the optimal effective
potential ${v}^{xOEP}$, therefore the name optimized effective potential
method.  The optimized effective potential ${v}^{xOEP}$ can always be
expressed as
\begin{eqnarray}
{v}^{xOEP}({\bf r}) &=& {v}_{ext}({\bf r}) + {v}_{H}({\bf r}) 
+ {v}_{x}({\bf r}) \,.
\label{OEP-potential}
\end{eqnarray}
with the Hartree potential given as the Coulomb potential of the electron
density generated by the orbitals. As shown in Refs. [\onlinecite{sharp53}]
and [\onlinecite{talman76}] the optimized effective potential ${v}^{xOEP}$ is
obtained if the exchange potential potential ${v}_{x}$ of Eq.\
(\ref{OEP-potential}) obeys the xOEP or EXX equation (\ref{EXX}).

Alternatively the xOEP or EXX equation (\ref{EXX}) can be derived within an
exact exchange-only KS framework. The Hamiltonian operator $\hat{H}^{xKS}$ of
the exact exchange-only KS equation is given by Eq.\ (\ref{KS-Hamiltonian})
with the effective KS potential
\begin{eqnarray}
{v}_s({\bf r}) &=& {v}_{ext}({\bf r}) + {v}_{H}({\bf r}) 
+ {v}_{x}({\bf r}) \,.
\label{KS-potential}
\end{eqnarray}
The KS exchange potential in Eq.\ (\ref{KS-potential}) is defined as the
functional derivative of the exchange energy
\begin{eqnarray}
E_x &=& - \;
\sum_i^{occ.}\sum_j^{occ.} \int \! d{\bf r} \int \! d{\bf r}' \;
\frac{\phi_i({\bf r}')\phi_j({\bf r}')\phi_j({\bf r})\phi_i({\bf r})}
{|{\bf r}-{\bf r}'|}
\label{Ex}
\end{eqnarray}
with respect to the electron density $\rho$, i.e, as
\begin{eqnarray}
v_x({\bf r}) &=& \frac{\delta E_x}{\delta \rho({\bf r})} \,.
\label{vx}
\end{eqnarray}
Following Ref. [\onlinecite{gorling95b,gorling05}] we now exploit that
according to the Hohenberg-Kohn theorem there exists a one-to-one mapping
between effective potentials ${v}_s$ and resulting electron densities $\rho$.
Therefore all quantities that are functionals of the electron density, here in
particular the exchange energy, can be simultaneously considered as
functionals of the effective potential $v_s$. Taking the functional derivative
$\delta E_x/\delta v_s({\bf r})$ of the exchange energy with respect to the
effective potential $v_s$ in two different ways with the help of the chain
rule yields
\begin{eqnarray}
\int \! d{\bf r}' \; \frac{\delta E_x}{\delta \rho({\bf r}')} \;
\frac{\delta \rho({\bf r}')}{\delta v_s({\bf r})}  
&=& 
\sum_a^{occ.} \int \! d{\bf r}' \;
\frac{\delta E_x}{\delta \phi_a({\bf r}')} \;
\frac{\delta \phi_a({\bf r}')}{\delta v_s({\bf r})}  \,.
\label{chain-rule}
\end{eqnarray}
The functional derivative $\delta\rho({\bf r})/\delta v_s({\bf r}')$ equals
the response function (\ref{Xs}) and the right hand side of Eq.\
(\ref{chain-rule}) equals the right hand side of the xOEP or EXX equation
(\ref{EXX}). Furthermore the response function $X_s$ is symmetric in its
arguments for real valued orbitals.  Therefore Eq.\ (\ref{chain-rule}) is
identical to the OEP or EXX equation (\ref{EXX}).  This shows that the
exchange potentials arising in the xOEP and the exact exchange-only KS schemes
and subsequently the xOEP and the exact exchange-only KS schemes itself are
identical. The xOEP or EXX equation can be derived in various ways within a KS
framework \cite{gorling05}.  A crucial point, however, is that all derivations
within a KS framework rely on real space representations in the sense that
functional derivatives are taken within real space because the KS exchange
potential is defined in real space as the functional derivative $\delta
E_x/\delta\rho({\bf r})$. Thus the above conclusion that the xOEP and the
exact exchange-only KS schemes are equivalent holds only in real space, i.e.,
if all quantities are respresented in real space.  Calculations, however, are
usually carried out in basis sets and we will show next that in this case an
xOEP and an exact exchange-only KS scheme, in general, are not equivalent.

The xOEP or EXX equation (\ref{EXX}) turns into the matrix equation
\begin{eqnarray}
{\bf X}_s \; {\bf v}_x  &=& {\bf t}
\label{EXX-basis}
\end{eqnarray}
with matrix and vector elements elements
\begin{eqnarray}
{X}_{s,k\ell} &=& 
4 \; \sum_i^{occ.}\sum_a^{unocc.} \;
\frac{\langle \phi_i |f_{k}| \phi_a \rangle 
      \langle \phi_a |f_{\ell}| \phi_i \rangle}
{\varepsilon_a- \varepsilon_s}\,,
\label{Xs-elements}
\end{eqnarray}
\begin{eqnarray}
{v}_{x, k} &=& \int \! d{\bf r} \; f_{k}({\bf r}) \, v_x({\bf r}) \,,
\label{vx-elements}
\end{eqnarray}
and
\begin{eqnarray}
{\bf t}_{k}
&=& 
4 \sum_i^{occ.}\sum_a^{unocc.} \;
\frac{\langle \phi_i | f_{k} | \phi_a \rangle
      \langle \phi_a | v_x^{NL} | \phi_i \rangle}
{\varepsilon_i - \varepsilon_a} \,.
\label{t-elements}
\end{eqnarray}
if an auxiliary basis set $\lbrace f_k \rbrace$ is introduced to represent the
response function, the exchange potential, and the right hand side of the EXX
equation (\ref{EXX}). For simplicity we assume at this point that the
auxiliary basis set is an orthonormal basis set. This is actually the case for
plane wave basis sets but not for Gaussian basis sets. However, without
changing the following arguments we can assume that we have orthonormalized
any auxiliary Gaussian basis set.

As long as the orbitals are represented in real space there is an infinite
number of them and the summations over unoccupied orbitals in the response
function (\ref{Xs}) and the right hand side of the xOEP or EXX equation
(\ref{EXX}) remains infinite and complete. For simplicity we assume that the
considered electron system is either periodic and thus exhibits periodic
boundary conditions or, in case of a finite system, is enclosed in a large but
finite box with an infinite external potential outside the box. Then the
number of orbitals is infinite but countable. As long as all orbitals are
taken into account in the summation over unoccupied orbitals, the basis set
representation of the exchange potential resulting from the basis set xOEP or
EXX equation (\ref{EXX-basis}) becomes the more accurate the larger the
auxiliary basis set and converges against the real space representation of the
exchange potential and can be interpreted both as exact exchange-only KS or
xOEP exchange potential.

This changes dramatically if the orbitals are represented in a finite orbital
basis set. Then, provided a reasonable orbital basis set is chosen, the
occupied and the energetically low unoccupied orbitals are well represented.
Most of the energetically higher unoccupied orbitals, however, are not
represented at all simply because a finite orbital basis set can not give rise
to an infinite number of unoccupied orbitals.  Moreover, the energetically
higher orbitals arising in a finite orbital basis set are quite poor
representations of true unoccupied orbitals. Let us now concentrate on the
representation of the response function. The integrals $\langle \phi_a |f_{k}|
\phi_i \rangle$ occuring in the matrix elements (\ref{Xs-elements}) of the
response function contain the three functions $\phi_i$, $\phi_a$, and $f_{k}$.
The occupied orbitals $\phi_i$ have few nodes and thus are relatively smooth
functions. The energetically low lying unoccupied orbitals still are
relatively smooth, the higher ones however, with an increasing number of nodes
and with increasing kinetic energy become more and more rapidly oscillating.
For smooth auxiliary basis functions $f_{k}$ the integrals $\langle \phi_a
|f_{k}| \phi_i \rangle$ approach zero if they contain an energetically high
unoccupied orbital $\phi_a$ because the product of the smooth functions
$f_{k}$ and $\phi_i$ again is a smooth function and the integral of this
smooth product with a rapidly oscillatory unoccupied orbital $\phi_a$ is zero
due to the fact that any integral of a smooth with a rapidly oscillating
function vanishes. This means that for matrix elements ${X}_{s,k\ell}$ of the
response function with two sufficiently smooth functions $f_{k}$ and
$f_{\ell}$ the summation over unoccupied orbitals in Eq.\ (\ref{Xs-elements})
can be restricted to unoccupied orbitals $\phi_a$ below a certain energy
depending on the smoothness of the involved auxiliary basis functions $f_{k}$
and $\chi_{\ell}$. For sufficiently smooth functions $f_{k}$ and $\chi_{\ell}$
the contributing unoccupied orbitals $\phi_a$ thus are well represented in a
finite orbital basis set. Therefore the matrix elements ${X}_{s,k\ell}$ of the
response functions are correct for indices $k$ and $\ell$ referring to
sufficiently smooth auxiliary basis functions.  For a more rapidly oscillating
auxiliary basis function nonvanishing matrix elements $\langle \phi_a |f_{k}|
\phi_i \rangle$ with energetically high unoccupied orbitals $\phi_a$ occur.
The energetically high unoccupied orbitals $\phi_a$, however, are poorly
described in the finite orbital basis set and moreover there are too few of
them. Therefore the matrix elements ${X}_{s,k\ell}$ of the response functions
turn out to be wrong if at least one index refers to a more rapidly
oscillating auxiliary basis functions. Indeed, if an auxilliary function $f_k$
oscillates much more rapidly than the energetically highest unoccupied
orbitals $\phi_a$ obtained for a given orbital basis set then all matrix
elements $\langle \phi_a |f_{k}| \phi_i \rangle$ and thus all corresponding
elements ${X}_{s,k\ell}$ of the response function is erroneously zero.

For a given auxiliary basis set, according to the above argument, a
representation of the response function is correct only if the orbital basis
set is balanced to the auxiliary basis set in the sense that it describes well
unoccupied orbitals up to a sufficiently high energy. Otherwise an incorrect
representation of the response function is obtained. The matrix representation
of the response function like the response function itself is negative
semidefinite. This is easily seen if a matrix element of the type $\langle f |
X_s| f \rangle$ for an arbitrary function $f$ is considered. Such a matrix
element is obtained by summing up the contributions occuring in the summation
over occupied and unoccupied orbitals in Eq.\ (\ref{Xs}). Each single
contribution and thus also the complete sum is nonpositive. Therefore an
insufficient orbital basis set leading to too few energetically high
unoccupied orbitals results in eigenvalues of the response matrix that have a
too small magnitude. Solutions of the matrix equation (\ref{EXX-basis}) are
given by the product of the inverse of the response matrix with the right hand
side of the equation, i.e., by ${\bf X}_s^{-1} \, {\bf t}$. If ${\bf X}_s$
contains eigenvalues that are too small then the corresponding eigenvectors
contribute with a too large magnitude to the solution of equation
(\ref{EXX-basis}). The eigenvectors with too small eigenvalues correspond to
rapidly oscillatory functions.  Therefore the resulting exchange potential
exhibits rapidly oscillatory features. This is exactly what is observed in the
xOEP scheme of Staroverov, Scuseria, and Davidson \cite{staroverov06}.  If the
response matrix even contains eigenvectors with eigenvalues that are
erroneously zero then an infinite number of solutions arise of the matrix
equation (\ref{EXX-basis}) corresponding to an infinite number of exchange
potentials, which yield, within the finite basis set, the same KS orbitals.

Therefore if the auxilliary and the orbital basis sets are chosen unbalanced,
e.g., if one chooses a too small orbital basis set for a given auxiliary basis
set or a too large auxiliary basis set for a given orbital basis set, then the
resulting response matrix ${\bf X}_s$ is corrupted and no longer represents a
proper representation of the response function in real space. In this case the
xOEP scheme no longer represents an exact exchange KS scheme and the resulting
exchange potential is unphysical and no longer represents the KS exchange
potential. However, even in this case the xOEP scheme still is a proper
optimized potential scheme in the sense that it yields a linear combination of
auxiliary basis functions that results in the lowest total energy for this
orbital basis set that can be obtained if the exchange potential shall be a
linear combination of the auxiliary basis functions. While the resulting
exchange potential is unphysical and does not resemble the KS exchange
potential it obeys the above requirement of the xOEP scheme. The reason is
that the arguments used for the xOEP derivation of the real space EXX or xOEP
equation can also be used if orbital and auxiliary basis sets are introduced
whereas no analogue to the DFT derivation exists anymore in this case.

\section{\label{examples}Examples}
We now illustrate the arguments of the previous two Sections by specific
examples. These examples also demonstrate that an auxiliary basis set that
consists of all products of occupied and unoccupied orbitals is not balanced
to the corresponding orbital basis set in the sense that a correct
representation of the response function and a proper KS exchange potential can
not be obtained for such an auxiliary basis set.  Firstly a system of
electrons in a box with periodic boundary conditions and an external potential
equal to a constant is considered. The box shall be defined by corresponding
unit cell vectors ${\bf a}_i$ with $i=1,2,3$. If the box, i.e., the unit cell
vectors, become infinitely large then the system turns into an homogeneous
electron gas. The KS eigenstates $\phi_{\bf G}$ of such a system are
determined by symmetry and are simple plane waves $\chi_{\bf G}$ as they are
given in Eq.\ (\ref{plane-waves}). All plane waves with ${\bf G}$ vectors of a
length smaller than some given constant G$_{F}$, i.e., with $|{\bf G}| \le
G_{F}$, shall represent occupied KS orbitals, all plane waves with $|{\bf G}|
> G_{F}$ represent unoccupied KS orbitals. The maximal length ${G_F}$ of the
vectors ${\bf G}$ of the occupied orbitals determines the Fermi level. For the
orbital basis set as well as for the auxiliary basis set we chose plane waves,
$\chi_{\bf G}$ and $f_{\bf G}$, respectively, again as given in Eq.\
(\ref{plane-waves}).  Thus, for the considered system, arises the special case
that each orbital basis function $\chi_{\bf G}$ represents a KS orbital
$\phi_{\bf G}$. Obviously, the cutoff $G_{cut}$ of the orbital basis set has
to be chosen equal to or larger than $G_{F}$.

The matrix representation ${\bf X}_s$ of the response function in the
considered case is diagonal with diagonal elements
\begin{eqnarray}
{X}_{s,{\bf G}{\bf G}} &=& 
4 \; \sum_{|{\bf G}'| \le G_F} \;
\frac{\langle \phi_{{\bf G}'} f_{{\bf G}}| 
                        \phi_{{\bf G}'+{\bf G}} \rangle 
      \langle \phi_{{\bf G}'+{\bf G}} |f_{{\bf G}} \, \phi_{{\bf G}'} \rangle}
{(|{\bf G}'|^2-|{\bf G}'+{\bf G}|^2)/2} \nonumber \\
&=& 
\frac{8}{V} \; \sum_{|{\bf G}'| \le G_F} \;
\frac{1}{(|{\bf G}'|^2-|{\bf G}'+{\bf G}|^2)} \,.
\label{XsGG}
\end{eqnarray}
The auxiliary basis set shall be characterized by the cutoff radius
$G_{cut}^{aux}$, i.e., the auxiliary basis set shall consist of all plane
waves $f_{\bf G}$ with $0 < |{\bf G}| \le G_{cut}^{aux}$.  Note that the
auxiliary function with ${\bf G} = {\bf 0}$ that equals a constant function
has to be excluded from the auxiliary basis set because the xOEP or EXX
equation in agreement with the basic formalism determines the exchange
potential only up to an additive constant. A constant function would be an
eigenfunction of the reponse function with zero eigenvalue.  Now three cases
can be distinguished: (i) If $G_{cut}^{aux} \le G_{cut}- G_F $ then the
corresponding matrix elements ${X}_{s,{\bf G}{\bf G}}$ of the reponse function
are obtained with their correct value in a basis set calculation with an
orbital basis set characterized by the cutoff radius $G_{cut}$ because all
unoccupied orbitals $\phi_{{\bf G}'+{\bf G}}$ occuring in the summation in
Eq.\ \ref{XsGG} can be represented by the orbital basis set.  (ii) If $G_{cut}
- G_F < G_{cut}^{aux} \le G_{cut} + G_F$ then for the matrix elements
${X}_{s,{\bf G}{\bf G}}$ with $G_{cut} - G_F < |{\bf G}| \le G_{cut} + G_F$
incorrect values are obtained because some of the unoccupied orbitals
$\phi_{{\bf G}'+{\bf G}}$ occuring for these matrix elements in the sum in
Eq.\ \ref{XsGG} can not be represented in the orbital basis set and therefore
are not taken into account. Because all terms in the sum in Eq.\ \ref{XsGG}
have the same sign the magnitudes of the resulting matrix elements
${X}_{s,{\bf G}{\bf G}}$ are too small.  (iii) If $G_{cut} + G_F <
G_{cut}^{aux}$ then the resulting ${\bf X}_{s}$ not only contains elements
with a too small magnitude but additionally all matrix elements ${X}_{s,{\bf
    G}{\bf G}}$ with $G_{cut} + G_F < |{\bf G}|$ are erroneously zero because
all of the unoccupied orbitals $\phi_{{\bf G}'+{\bf G}}$ occuring in the
summation in Eq.\ \ref{XsGG} can not be represented in the orbital basis set
and therefore are not taken into account.

If the auxiliary basis set is chosen to be the space spanned by all products
of occupied and unoccupied orbitals then it consists of all plane waves
$f_{\bf G}$ with $0 < |{\bf G}| \le G_{cut}+ G_F$, i.e., $G_{cut}^{aux} =
G_{cut}+ G_F$. Thus the auxiliary basis set is chosen according to the above
cases (ii). Therefore some of the resulting matrix elements ${X}_{s,{\bf
    G}{\bf G}}$ of the reponse function are incorrect. This demonstrates that
an auxiliary basis set given by all products of occupied and unoccupied
orbitals is not balanced with the corresponding orbital basis set.

The considered system is special in that the right hand side of the xOEP or
EXX matrix equation is zero due to the translational symmetry. Therefore also
the resulting exchange potential is zero or more precisely equals an arbitrary
constant. If the auxiliary basis set is chosen according to the above cases
(i) and (ii) then a basis set calculation yields the correct exchange
potential, i.e., zero or a constant.  If the auxiliary basis set contains
functions according to the above case (iii), however, then the xOEP or EXX
matrix equation erroneously has an infinite number of solutions that equal a
constant plus an arbitrary contribution of auxiliary basis functions with
$G_{cut} + G_F < |{\bf G}|$.  The reason why the correct exchange potential is
obtained for an auxiliary basis set chosen according to the above case (ii)
despite the fact that in this case the response function is already corrupted
is that for the special system considered here the right hand side of the xOEP
or EXX matrix equation is zero. Therefore any values for the diagonal elements
${X}_{s,{\bf G}{\bf G}}$ that differ from zero lead to the correct result.
However, in general the right hand side of xOEP or EXX matrix equation is not
equal to zero and then a response matrix with eigenvalues with erroneously too
small magnitudes leads to a wrong exchange potential that exhibits too large
contributions from those linear combinations of auxiliary basis functions that
correspond to the too small eigenvalues of the response matrix. This is
demonstrated in the following example.

We consider plane wave xOEP calculations for bulk silicon carried out with the
method of Ref. [\onlinecite{stadele97}]. The integrable singularity occuring
in HF and xOEP exchange energies in plane wave treatments of solids is taken
into account according to Ref. [\onlinecite{carrier06}].  The lattice constant
was set to the experimental value of 5.4307\AA. The set of used {\bf k}-points
was chosen as a uniform $4\times4\times4$ mesh covering the first Brillouin
zone.  In all calculations, all unoccupied orbitals resulting for a given
orbital basis set were taken into account for the construction of the response
function and the right hand side of the xOEP equation. EXX pseudopotentials
\cite{engel01,engel01b} with angular momenta $l=0,1,2$ and cutoff radii, in
atomic units, of $r_{c,l=0}^{Si}\!=\!1.8$, $r_{c,l=1}^{Si}\!=\!2.0$, and
$r_{c,l=2}^{Si}\!=\!2.0$ were employed. The pseudopotential with $l=1$ was
chosen as local pseudopotential.

\begin{figure}[h]
\includegraphics*[width=11.0cm]{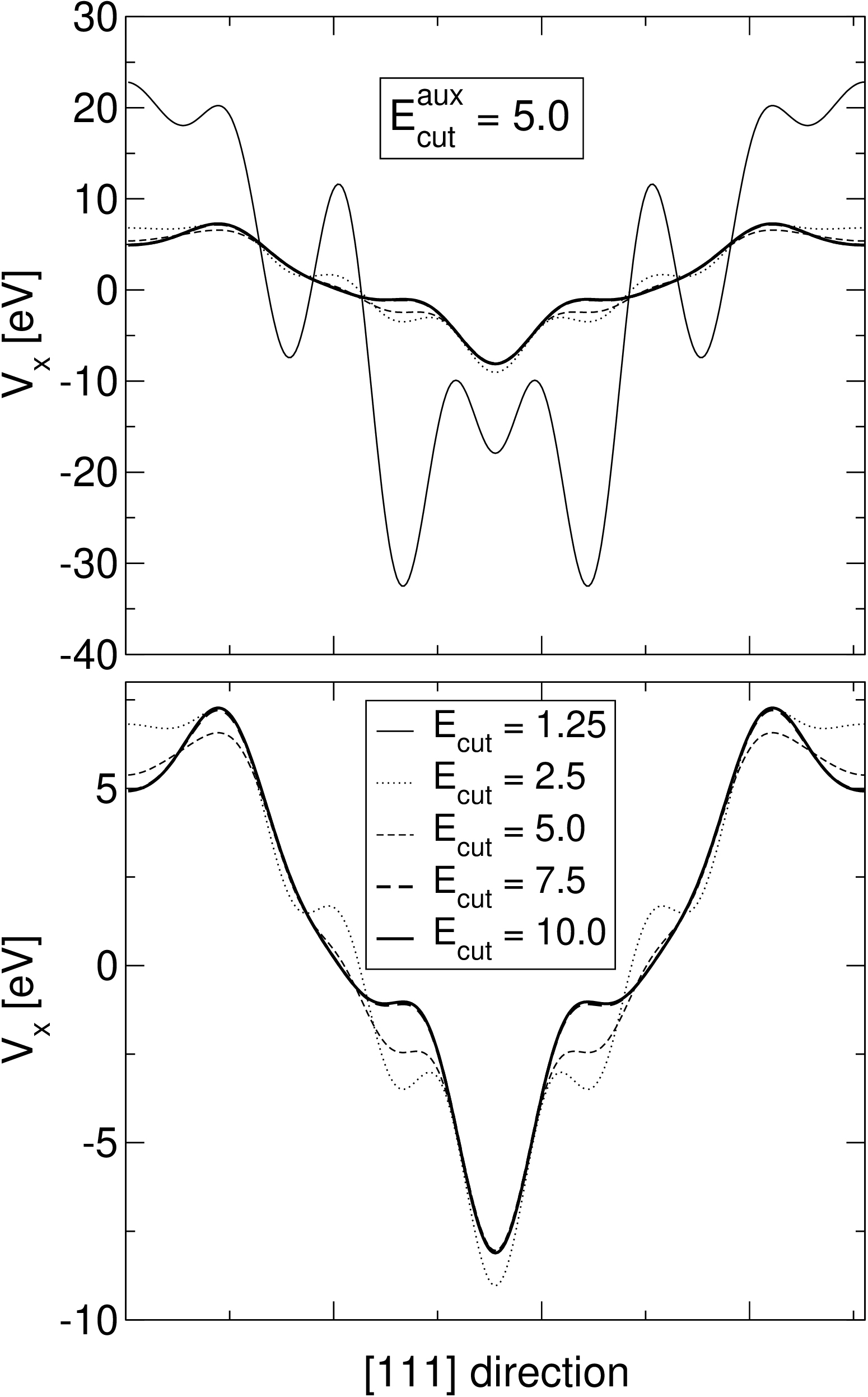}
\caption
{xOEP exchange potential along the silcion-silicon bond axis, i.e., the unit
cell's diagonal, 
for an auxiliary basis set cutoff $E_{cut}^{aux} = 5.0 \, a.u.$
and different orbital basis set cutoffs $E_{cut}$. The upper and lower panels
differ in the energy scale. The curve for $E_{cut} = 1.25 \, a.u.$ is only
displayed in the upper panel.}
\label{fig.chicut5}
\end{figure}

\begin{figure}[h]
\includegraphics*[width=11.0cm]{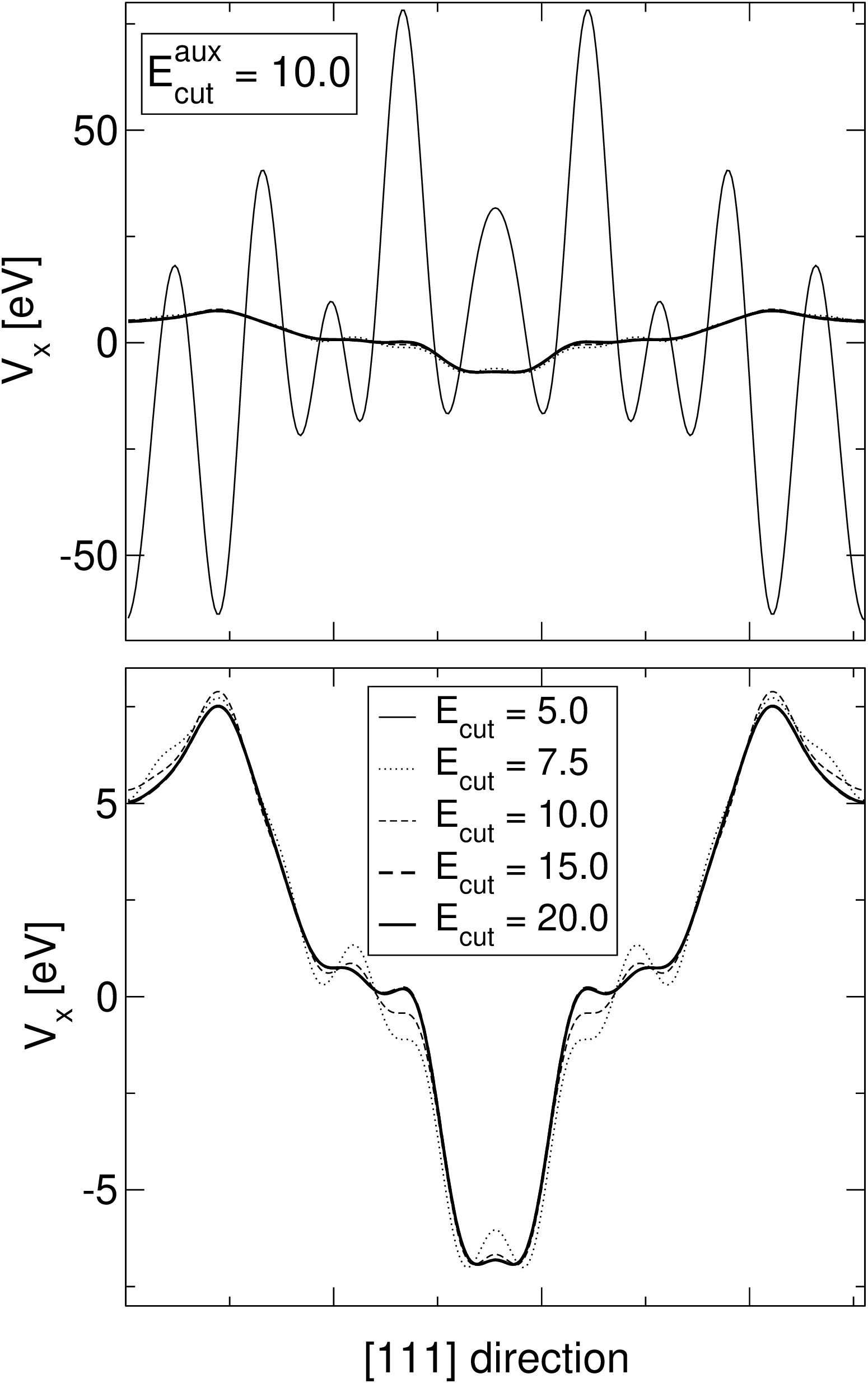}
\caption
{xOEP exchange potential along the silcion-silicon bond axis, i.e., the unit
cell's diagonal, 
for an auxiliary basis set cutoff $E_{cut}^{aux} = 10.0 \, a.u.$
and different orbital basis set cutoffs $E_{cut}$. The upper and lower panels
differ in the energy scale. The curve for $E_{cut} = 5.0 \, a.u.$ is only
displayed in the upper panel.}
\label{fig.chicut10}
\end{figure}

Figs.\ \ref{fig.chicut5} and \ref{fig.chicut10} display xOEP exchange
potentials along the silcion-silicon bond axis, i.e., the unit cell's
diagonal, for auxiliary basis set cutoffs $E_{cut}^{aux}$ of 5.0 and 10.0 a.u.
($G_{cut}^{aux}$ of 3.2 and 4.5 a.u.), and for various different orbital basis
set cutoffs $E_{cut}$.  Note that in figures and tables instead of the cutoffs
$G_{cut}$ and $G_{cut}^{aux}$ that refer to the length of the reciprocal
lattice vectors of the plane waves the corresponding energy cutoffs $E_{cut} =
\frac{1}{2}G_{cut}^2$ and $E_{cut}^{aux} = \frac{1}{2}(G_{cut}^{aux})^2$ are
displayed.  Fig.\ \ref{fig.chicut5} shows that the combination of an auxiliary
basis set with cutoff $E_{cut}^{aux} = 5.0$ ($G_{cut}^{aux} = 3.2$) with a
orbital basis set with cutoff $E_{cut} = 1.25$ ($G_{cut} = 1.6$) leads to a
highly oscillating unphysical exchange potential.  The cutoff of the auxiliary
basis set in the considered case is about twice as large as the cutoff of the
orbital basis. This means that the space spanned by the auxiliary basis is the
same as that of all products of occupied and unoccupied orbitals. In this case
the matrix representing the response function is corrupted and the resulting
exchange potential turns out to be unphysical. With increasing cutoff
$E_{cut}$ of the orbital basis set the xOEP exchange potentials converge
towards the physical KS exchange potential, more precisely towards the
representation of the physical KS exchange potential in an auxiliary basis set
with cutoff $E_{cut}^{aux} = 5.0$ ($G_{cut}^{aux} = 3.2$).  If the cutoff
$E_{cut}$ of the orbital basis set is about 1.5 times as large as the cutoff
of the auxiliary basis set $E_{cut}^{aux}$, i.e., equals 7.5 ($G_{cut}= 3.9$),
then the exchange potential is converged. A further increase of $E_{cut}$ to
$E_{cut}= 10.0$ ($G_{cut}= 4.5$) leads to an exchange potential that is
indistinguishable from that for $E_{cut}= 7.5$ ($G_{cut}= 3.9$) on the scale
of Fig.\ \ref{fig.chicut5}.  Fig.\ \ref{fig.chicut10} gives an analogous
picture for a cutoff of the auxiliary basis set of $E_{cut}^{aux} = 10.0$
($G_{cut}^{aux} = 4.5$).  Again, if the space spanned by the auxiliary basis
set equals that of the product of occupied and unoccupied orbitals, curve for
$E_{cut}= 5.0$ ($G_{cut}= 3.2$), an highly oscillating unphysical exchange
potential is obtained. If $E_{cut} \approx 1.5 E_{cut}^{aux}$ then the
exchange potential is converged towards the representation of the physical KS
exchange potential in an auxiliary basis set with cutoff $E_{cut}^{aux} =
10.0$ ($G_{cut}^{aux} = 4.5$).

This demonstrates the point that the xOEP scheme only represents a KS scheme
if the orbital basis set is balanced to the auxiliary basis set. In the case
of a plane wave basis set this requires the energy cutoff $E_{cut}$ of the
orbital basis set to be about 1.5 times larger than the energy cutoff
$E_{cut}^{aux}$ of the auxiliary basis set.

Table \ref{tab.energies} lists for a number of orbitals basis set cutoffs
$E_{cut}$ exchange and ground state energies for series of auxiliary basis set
cutoffs $E_{cut}^{aux}$. Table \ref{tab.energies} shows that the ground state
energies for a given $E_{cut}$ always decrease with increasing $E_{cut}^{aux}$
even if the values of $E_{cut}^{aux}$ is that large that the resulting
exchange potential is unphysical. This demonstrates that the xOEP scheme
remains well-defined even if unbalanced basis sets are used.  In this case,
however, the xOEP scheme no longer represents a KS method and the resulting
exchange potential is unphysical and does not represent the KS exchange
potential.  Table \ref{tab.energies} also lists the differences of the xOEP
and HF ground state energies and shows that the xOEP energy does not converge
to the HF energy. In the combinations $E_{cut}= 2.5 / E_{cut}^{aux}= 10.0$,
$E_{cut}= 5.0 / E_{cut}^{aux}= 20.0$, and $E_{cut}= 7.5 / E_{cut}^{aux}= 29.9
$ the space spanned by the auxiliary basis set roughly equals that of the
product of occupied and unoccupied orbitals. The ground state xOEP energies in
these cases is { de facto} the lowest that can be achieved by the xOEP method
for the given orbital basis set. The fact that this energy is higher than the
HF total energy shows that the xOEP energy does not reach the HF ground state
energy if the products of occupied and unoccupied orbitals become linearly
dependent as it is usually the case in plane wave calculations and as it is
the case in the presented calculations.

\section{\label{summary}Summary}
We have given arguments leading to the conclusion that exchange-only optimized
potential (xOEP) methods, with finite basis sets, cannot in general yield the
Hartree-Fock (HF) ground state energy, but a ground state energy that is
higher. This holds true even if the exchange potential that is optimized in
xOEP schemes is expanded in an arbitrarily large auxiliary basis set. The HF
ground state energy can only be obtained via an xOEP scheme in the special
case that all products of occupied and unoccupied orbitals emerging for the
orbital basis set are linearly independent from each other. In this case,
however, exchange potentials leading to the HF ground state energy exhibit
unphysical oscillations and do not represent Kohn-Sham (KS) exchange
potentials. These findings solve the seemingly paradoxical results of
Staroverov, Scuseria and Davidson \cite{staroverov06} that certain finite
basis set xOEP calculations lead to the HF ground state energy despite the
fact that it was shown \cite{ivanov03} that within a real space representation
(complete basis set) the xOEP ground state energy is always higher than the HF
energy. A key point is that the orbital products of a complete basis are
linearly dependent.

Moreover, whether or not the products of occupied and unoccupied orbitals are
linearly independent, we have shown that basis set xOEP methods only represent
exchange-only (EXX) KS methods, i.e., proper density-functional methods, if
the orbital basis set and the auxiliary basis set representing the exchange
potential are balanced to each other, i.e., if the orbital basis set is
comprehensive enough for a given auxiliary basis set. Otherwise xOEP schemes
do not represent EXX KS methods.  We have found that auxiliary basis sets that
consist of all products of occupied and unoccupied orbitals are not balanced
to the corresponding orbital basis set. The xOEP method, even in cases of
unbalanced orbital and auxiliary basis sets, works properly in the sense that
it determines among all exchange potentials that can be represented by the
auxiliary basis set the one that yields the lowest ground state energy.
However, in these cases the resulting exchange potential is unphysical and
does not represent a KS exchange potential. Therefore the xOEP method is of
little practical use in those cases for which it does not represent a EXX KS
method. Remember that, at present, the main reason to carry out xOEP methods
in most cases is to obtain a qualitatively correct KS one-particle spectrum,
either for the purposes of interpretation or as input for other approaches
like time-dependent density-functional methods. However, the unphysical
oscillations of the exchange-potential of xOEP schemes with unbalanced basis
sets affect the unoccupied orbitals and eigenvalues. Another reason to carry
out xOEP methods that represent EXX KS methods is that the latter may be
combined with new, possibly orbital-dependent, correlation functionals to
arrive at a new generation of density-functional methods. Also in this case it
is important that the xOEP methods represents proper KS methods.

A balancing of auxiliary and orbital basis sets is straightforward for plane
wave basis sets. In this case xOEP schemes are proper EXX KS methods if the
energy cutoff for the orbital basis set set is about 1.5 times as large as
that of the auxiliary basis set. This as well as other results of this work
were illustrated with plane wave calculations for bulk silicon.  For Gaussian
basis sets on the other hand, a proper generally applicable and reasonably
simple balancing scheme of orbital and auxiliary basis sets is so far not
available despite much efforts
\cite{gorling99,ivanov99,hamel01,veseth01,hirata01,yang02}.  Therefore
effective exact exchange-only methods like the KLI \cite{krieger92}, the
'localized Hartree-Fock' \cite{dellasala01}, the equivalent 'common energy
denominator approximation' method \cite{gritsenko01}, or the closely related
very recent method of Ref. \onlinecite{staroverov06b}, are in use as
numerically stable alternatives that yield results very cose to those of full
EXX KS methods.

{\renewcommand{\baselinestretch}{1.1}

\begin{table}[h]
\caption{        xOEP exchange and ground state energies 
$E_\mathrm{x}^\mathrm{xOEP}$ and $E^\mathrm{xOEP}$, respectively,
 and the difference between
HF and xOEP ground state energy for various combinations of and orbital 
auxiliary basis sets, characterized by energy cutoffs $E_{cut}$ and
$E_{cut}^{aux}$, respectively. ($N$ and $M^\mathrm{aux}$ denote the
 corresponding number of basis functions.) All quantities are given in a.u.
\label{tab:energies}}
\begin{ruledtabular}
\begin{tabular}{lrrccc}
$E_{cut}$ / $N$ & $E_{cut}^{aux}$ & $M^\mathrm{aux}$ & 
                        \multicolumn{1}{c}{$E_{x}^{xOEP}$} & 
                        \multicolumn{1}{c}{$E^{xOEP}$} & 
                        \multicolumn{1}{c}{$E^{HF}$-E$^{xOEP}$} \\
\hline
 2.5 / 59  &  2.5  &   59 &   -2.1423 &     -7.4028 &  0.0054 \\      
           &  5.0  &  137 &   -2.1434 &     -7.4033 &  0.0050 \\      
           &  6.0  &  181 &   -2.1463 &     -7.4043 &  0.0039 \\      
           &  7.4  &  259 &   -2.1474 &     -7.4051 &  0.0031 \\      
           & 10.0  &  411 &   -2.1479 &     -7.4053 &  0.0030 \\ [2ex]
 5.0 / 150 &  2.5  &   59 &   -2.1451 &     -7.5061 &  0.0077 \\      
           &  5.0  &  137 &   -2.1460 &     -7.5065 &  0.0073 \\      
           &  7.4  &  259 &   -2.1468 &     -7.5069 &  0.0070 \\      
           & 10.0  &  411 &   -2.1481 &     -7.5076 &  0.0062 \\      
           & 14.9  &  725 &   -2.1501 &     -7.5087 &  0.0051 \\      
           & 20.0  & 1139 &   -2.1502 &     -7.5088 &  0.0050 \\ [2ex]
 7.5 / 274 &  2.5  &   59 &   -2.1482 &     -7.5269 &  0.0080 \\      
           &  5.0  &  137 &   -2.1487 &     -7.5272 &  0.0078 \\      
           &  7.4  &  259 &   -2.1494 &     -7.5274 &  0.0075 \\      
           & 10.0  &  411 &   -2.1495 &     -7.5275 &  0.0075 \\      
           & 14.9  &  725 &   -2.1520 &     -7.5286 &  0.0063 \\
           & 24.9  & 1639 &   -2.1539 &     -7.5296 &  0.0053 \\
           & 29.9  & 2085 &   -2.1540 &     -7.5297 &  0.0053 \\ [2ex]
 10.0 / 415&  2.5  &   59 &   -2.1489 &     -7.5287 &  0.0081 \\      
           &  5.0  &  137 &   -2.1494 &     -7.5290 &  0.0078 \\      
           &  7.4  &  259 &   -2.1500 &     -7.5292 &  0.0076 \\      
           & 10.0  &  411 &   -2.1501 &     -7.5292 &  0.0076 \\      
           & 14.9  &  725 &   -2.1505 &     -7.5294 &  0.0074 \\      
           & 20.0  & 1139 &   -2.1511 &     -7.5296 &  0.0072 \\ 
\end{tabular}           
\end{ruledtabular}
\label{tab.energies}
\end{table}

} 

\section{\label{appendix}Appendix: Linear dependence of products of basis
                        functions of a complete basis}
Let $\{\phi_{k}(x)\}$ be a complete set of functions of a complex valued
variable $x$ such that any arbitrary square integrable function 
can be written as a linear combination of the functions in the complete set. 
We show that the set $\{\phi_{k}(x)\phi_{l}(x)\}$  is linearly dependent.

Using our complete sets, an arbitrary function $f(x,y)$ of two complex valued
variables $x$ and $y$ may be expanded in terms of $\{\phi_{k}(x)\}$ and
$\{\phi_{\ell}(y)\}$
\begin{equation}
f(x,y)=\sum_{k=1}^{\infty}
\sum_{\ell=1}^{\infty}b_{k,\ell}\,\phi_{\ell}(y)\phi_{k}(x)
\end{equation}
Set $y=x$ to get:
\begin{equation}
\label{eq:1}
f(x,x)=\sum_{k=1}^{\infty}\sum_{\ell=1}^{\infty}\,
b_{k,\ell}\,\phi_{\ell}(x)\phi_{k}(x)
\end{equation}

Now choose a function $f(x,x)$ and a $\phi_n(x)$ out of the set
$\{\phi_{k}(x)\}$ such that (i) $\lim_{x\to \infty}\frac{f(x,x)}{\phi_n(x)}=0$
and (ii) at least one $b_{k,\ell}\neq0$ when $\ell \neq n$ and $k\neq n$.
Since $\frac{f(x,x)}{\phi_n(x)}$ is just a function of $x$, we may expand it
in term of the $\{\phi_{k}(x)\}$:
\begin{equation}
\frac{f(x,x)}{\phi_n(x)}=\sum_{m=1}^{\infty}d_{m}\phi_{m}(x)
\end{equation}
Solving for $f(x,x)$,
\begin{equation}
\label{eq:2}
f(x,x)=\phi_n(x)\sum_{m=1}^{\infty}
d_{m}\phi_{m}(x)=\sum_{m=1}^{\infty}d_{m}\phi_{m}(x)\phi_n(x)
\end{equation}
and equating Eq. (\ref{eq:1}) with Eq. (\ref{eq:2}), we get
\begin{equation}
\sum_{m=1}^{\infty}d_{m}\phi_{m}(x)
\phi_n(x)=\sum_{k=1}^{\infty}\sum_{\ell=1}^{\infty}b_{k,\ell}\phi_{k}(x)
\phi_{\ell}(x)
\end{equation}
or by setting $k=m$,
\begin{equation}
\sum_{m=1}^{\infty}d_{m}
\phi_{m}(x)\phi_n(x)-\sum_{m=1}^{\infty}
\sum_{\ell=1}^{\infty}b_{m,\ell}\phi_{m}(x)\phi_{\ell}(x)
=0
\end{equation}
or 
\begin{equation}
\label{eq:4}
(d_{n}-b_{n,n})\phi_{n}(x)\phi_{n}(x)+
\sum_{\substack{j=1 \\ j\neq n}}^{\infty}(d_{j}-b_{j,n}- b_{n,j})
\phi_{n}(x)\phi_{j}(x)-\sum_{\substack{\ell=1 \\ \ell \neq n}}^{\infty}
\sum_{\substack{m=1 \\ m \neq n}}^{\infty}b_{m,\ell}\phi_{m}(x)\phi_{\ell}(x)=0
\,.
\end{equation}
Eq. (\ref{eq:4}) is a linear combination of a subset of
$\{\phi_{k}(x)\phi_{l}(x)\}$ broken up into disjoint components and equated to
zero. If a subset of a set is linearly dependent, then the set must also be
linearly dependent.  We show such a case by contradiction: According to Eq.
(\ref{eq:4}), for the subset $\{\phi_{k}(x)\phi_{l}(x)\}_{\subset}$ (appearing
in the equation) to be linearly independent, three conditions must be met:
\begin{enumerate}
\item  $d_{n}=b_{n,n}$
\item  $d_{j}-b_{j,n}-b_{n,j}=0$ ($\forall$ $j \in \mathbb{N}$ with $j \neq n$)
\item $b_{m,\ell}=0$ ($\forall$ $m,\ell \in \mathbb{N}$ 
with $m \neq n$ and $\ell \neq n$)
\end{enumerate}
But according to our condition on $f(x,x)$ there is at least one
$b_{m,\ell}\neq0$ with $m \neq n$ and $\ell \neq n$, which is a contradiction
to number three of our linear independence criteria.  Therefore
$\{\phi_{k}(x)\phi_{l}(x)\}_{\subset}$ must be linearly dependent by
contradiction, and therefore $\{\phi_{k}(x)\phi_{l}(x)\}$ for all $k,l\in
\mathbb{N}$ is linearly dependent because
$\{\phi_{k}(x)\phi_{l}(x)\}_{\subset}$ is linearly dependent.

One may take the result one step further to show with an induction argument
that for any complete set such as $\{\phi_{k}(x)\}$ the set defined by
$\{\prod_{i=1}^{N}\phi_{p_{i}}(x)\vert$ $N,i,p_{i}\in \mathbb{N}$\} is
complete and linearly dependent.


\begin{thebibliography}{40}
\bibitem{constrainedsearch} See the 'constrained search‘ discussion 
            in Ref [\onlinecite{ivanov03}], 
            which includes appropriate references on the approach in this
            context .
\end{thebibliography}
\end{document}